\documentclass[twocolumn,showpacs,preprintnumbers,amsmath,amssymb]{revtex4}
\usepackage{epsfig}
\usepackage{bbm}
\newcommand{\bm}[1]{ \mbox{\boldmath $#1$}  }
\newcommand{\ud}{\mathrm{d}}

\begin{document}

\title{Cluster sum rules for three-body systems with angular-momentum 
dependent interactions}

\author{R. de Diego  and E. Garrido}
\affiliation{Instituto de Estructura de la Materia, CSIC, Serrano 123, E-28006
Madrid, Spain}

\author{A.S. Jensen and D.V. Fedorov}
\affiliation{Department of Physics and Astronomy, University of Aarhus,
DK-8000 Aarhus C, Denmark}
\date{\today}

\begin{abstract}
We derive general expressions for non-energy weighted and
energy-weighted cluster sum rules for systems of three charged
particles.  The interferences between pairs of particles are found to
play a substantial role. The energy-weighted sum rule is usually
determined by the kinetic energy operator, but we demonstrate that it
has similar additional contributions from the angular momentum and
parity dependence of two- and three-body potentials frequently used in
three-body calculations.  The importance of the different
contributions is illustrated with the dipole excitations in
$^6$He. The results are compared with the available experimental data.
\end{abstract}


\maketitle

\section{Motivation}

The use of sum rules in quantum mechanics is well established and
abundantly applied for many different systems \cite{sha74,boh75}. The
prominent examples are the transitions from a given quantum state
induced by an electromagnetic multipole operator. For any multipole
operator acting on an initial state, the sum of all the related
transition probabilities multiplied by powers of the excitation energy
are completely determined by the properties of the initial state
\cite{boh79,lip89}.  

The sum rules exist in general for
any many-body quantum system. Of specific interest are those systems
where the constituents clusterize, such that the degrees of freedom
can be divided into the internal ones corresponding to each cluster,
and those associated with the relative motion of the clusters
\cite{jen04}. Then the different multipole operators can be decomposed into terms
depending on the intrinsic coordinates of each cluster, and an
additional term depending only on the relative coordinates of the
centers of mass of the clusters.  This operator structure then leaves
two sum rules showing the same decomposition, the sum rules associated
with each individual cluster (depending only on the properties of the
initial cluster state), plus the {\it cluster sum rule} (depending on
the properties of the few-body initial wave function).  Examples are
found in \cite{sag90,suz03}, where the dipole non-energy weighted and
dipole energy-weighted sum rules are obtained for many-body systems
clusterizing into a two-body system.

When a clusterized system can be properly described as a few-body
system where the internal cluster degrees of freedom are frozen, only
the cluster sum rules remain, corresponding to the much smaller
Hilbert space of ground and excited states of the relative cluster
motion.  This kind of few-body descriptions have been extensively used
in nuclear physics during the latest 10-15 years in connection with
halos and weakly bound states in general \cite{jen04}.  The most
interesting and frequently investigated of these systems are
approximated by a three-body structure.  Extensions to excited
three-body continuum states are now being pursued and attracting a lot
of attention \cite{myo01,des06,alv07,kan07,mas07}. To get accurate
three-body wave functions the Faddeev decomposition with different
Jacobi coordinates is employed in coordinate space computations
\cite{nie01}.  The unavoidable transformation from one set of Jacobi
coordinates to another complicates the structure of the cluster sum
rules, especially when more than one of the three particles is
charged.

The purpose of this work is to generalize the dipole two-body cluster
sum rule as in \cite{sag90,suz03} to three-body systems for any
multipolarity. Advanced three-body calculations employ partial wave
dependent pair interactions and state dependent three-body potentials,
and these complications must therefore also be taken into account in
derivations of the cluster sum rules.  Typically, the two-body
interactions are adjusted independently for each partial wave in order
to reproduce the available properties of the corresponding two-body
system, e.g. bound state and resonance energies, and phase shifts
\cite{gar02,gar04}.  These interactions are then essentially non-local 
through their angular momentum dependence.  Also, it is common to
introduce effective angular momentum and parity dependent three-body
forces for fine-tuning the crucial total energies of the three-body
states.  Since the multipole transition operators carry both angular
momentum and parity they do not commute, in general, with the angular
momentum dependent two- and three-body potentials.  Thus, the energy
weighted cluster sum rule should then be rederived including the
corresponding contributions in addition to the usual kinetic energy
term.  In all cases we must include contributions from the three
Faddeev components which are expressed in their respective Jacobi
coordinates.

In section \ref{sec2} we briefly introduce the coordinates used, and
summarize some important relations and definitions. In section
\ref{sec3} we derive the non-energy weighted sum rule. The energy-weighted
sum rule is obtained in section \ref{sec4}, which is divided into
three subsections corresponding to the contributions from the kinetic
energy operator, the partial wave dependent two-body potentials,
and the (total angular momentum) $J^\pi$-dependent three-body
forces. As an illustration, in section \ref{sec5} we investigate the
dipole excitations in $^6$He and compare with the available
experimental data. We close the paper with a short summary and the
conclusions. A few intermediate expressions obtained in the derivations 
have been collected in the appendix.

\section{The transition probability}
\label{sec2}

We assume three clusters with masses $m_i$ and charges $z_i$
($i$=1,2,3), described by coordinates $\bm{r}_i$, and with the
three-body center-of-mass at $\bm R$. The three sets of mass-scaled
Jacobi coordinates are $\{\bm{x}_i,\bm{y}_i\} \equiv
\{\rho,\alpha_i,\Omega_{x_i},\Omega_{y_i}\}$, where $\rho$ is the hyperradius
and $\{\alpha_i,\Omega_{x_i},\Omega_{y_i}\}$ are the five hyperangles
corresponding to the Jacobi set $i$, see for instance \cite{nie01}.
The connection between the Cartesian and the mass-scaled Jacobi
coordinates is given by
\begin{eqnarray}
 \bm{r}_j -\bm{r}_k&=& \sqrt{\frac{m_N(m_j+m_k)}{m_j m_k}} \bm{x}_i\;, \\
 \bm{r}_i - \bm{R}  &=& \sqrt{\frac{m_N}{m_i} \frac{(m_j+m_k)}{M}} \bm{y}_i
 \; ,\label{eq5}
\end{eqnarray}
where $M$=$m_i$+$m_j$+$m_k$ and $m_N$ is the normalization mass.  The
transformation between different sets of Jacobi coordinates are given
by
\begin{eqnarray}
\label{eq6}
 \bm{y}_i &=& -\bm{x}_k \sin \varphi_{ik} - \bm{y}_k\cos\varphi_{ik} \;,\\
 \tan \varphi_{ik} &=& (-1)^p \sqrt{\frac{m_j M}{m_i m_k}} \hspace*{1mm} 
(\mbox{if } i\neq k), \mbox{ and } \varphi_{ii}=\pi,
\label{eq7}
\end{eqnarray}
which formally amounts to a rotation depending on the mass ratios and
the parity $(-1)^p$ of the permutation $p$ of $\{i,j,k\}$.
Eqs.(\ref{eq6}) and (\ref{eq7}) lead to an important relation between
harmonic polynomials in different Jacobi coordinates, i.e.
\begin{eqnarray}
 y_i^{\lambda} Y_{\lambda, \mu}(\Omega_{y_i}) &=&
 \sum_{\ell=0}^{\lambda} (-1)^{\lambda} x_k^{\lambda-\ell}
 (\sin \varphi_{ik})^{\lambda-\ell} y_k^{\ell} (\cos \varphi_{ik})^{\ell} \nonumber \\
&& \hspace*{-2.5cm}
\sqrt{\frac{4\pi (2\lambda+1)!}{(2\ell+1)!(2\lambda-2\ell+1)!}}
\left[ Y_{\lambda-\ell}(\Omega_{x_k}) \otimes
 Y_{\ell}(\Omega_{y_k})\right]^{\lambda \mu}.
\label{eq8}
\end{eqnarray}

Let us consider the initial three-body state $|n_0J_0M_0\rangle$,
where $J_0$ is the total angular momentum with projection $M_0$. All
the other needed quantum numbers are collected into $n_0$. The excited
states $\left\{|nJM\rangle\right\}$ can be populated from the ground
state by the electric multipole operator
\begin{eqnarray}
\label{eq1}
 O_{\mu}^{\lambda} = \sum_{i=1}^3 z_i |\bm{r}_i - \bm{R}|^{\lambda}
 Y_{\lambda, \mu}(\Omega_{y_i}) \;,
\end{eqnarray}
where $i$ runs over the three clusters, or equivalently, over the three 
sets of Jacobi coordinates.

The transition probability corresponding to this electric multipole
operator is proportional to the ${\cal B}(E\lambda)$-value, i.e.
\begin{equation} \label{eq77}
   {\cal B}(E\lambda,n_0J_0 \rightarrow nJ) = \sum_{\mu M}
  |\langle nJM|O_{\mu}^{\lambda}|n_0J_0M_0\rangle|^2
\end{equation}
from which the $\lambda$-multipole strengths are defined as
\begin{equation}
S_m=\sum_{nJ} (E_{nJ} - E_{0})^m
   {\cal B}(E\lambda,n_0J_0 \rightarrow nJ),
\label{def}
\end{equation}
where $E_0$ is the energy of the initial state, and $E_{nJ}$ is the
energy of the excited state with angular momentum $J$ and additional
quantum numbers $n$.

The values of these multipole strengths, depending only on the
properties of the initial state, are known as the sum rules.  In this
work we are concentrating on the sum rules with $m$=0,1, also denoted
as non-energy weighted and energy-weighted sum rules, respectively.

For the dipole case ($\lambda$=1), after inclusion of Eq.(\ref{eq8}) 
into (\ref{eq1}), and using Eqs.(\ref{eq5}) and (\ref{eq7}), 
one can see that for three particles having equal value of the 
ratio $z_i/m_i$ the dipole operator is zero. This means that for the
particular case of three identical particles all the dipole strengths
$S_m$ in Eq.(\ref{def}) are zero.

\section{The non-energy weighted sum rule}
\label{sec3}

The sum over all transitions can be rewritten provided the
intermediate set of quantum numbers gives a complete description of
the (bound and continuum) final states, i.e. $\sum_{nJM} |nJM\rangle \langle
nJM|=\mathbbm{1}$. We then get:
\begin{eqnarray}
&& S_0 = \sum_{nJ} {\cal B}(E\lambda,n_0J_0 \rightarrow nJ) = \nonumber \\ &&\sum_{n J} \sum_{\mu M}
  \langle n_0J_0M_0 |O_{\mu}^{\lambda \dagger}|nJM\rangle \nonumber \langle nJM| O_{\mu}^{\lambda}|n_0J_0M_0 \rangle= \nonumber \\ &&\sum_{\mu}
  \langle n_0J_0M_0 | O_{\mu}^{\lambda \dagger} O_{\mu}^{\lambda}
 |n_0J_0M_0 \rangle \; ,
\label{eq3}
\end{eqnarray}
which for a given multipole operator is entirely determined by the
properties of the ground state $|n_0J_0M_0\rangle$.

The definition in Eq.(\ref{eq1}), together with (\ref{eq5}) and
(\ref{eq8}), permits expressing the operator $O_{\mu}^{\lambda \dagger}
O_{\mu}^{\lambda}$ in terms of a single set of Jacobi coordinates,
leading to
\begin{small}
\begin{eqnarray}
&& \hspace*{-5mm} S_0 = \sum_{i,k=1}^3 \sum_{\ell=0}^{\lambda}\sum_{\mu=-\lambda}^{\lambda} (-1)^\mu z_i z_k
 \sqrt{\frac{4\pi(2\lambda+1)!(2\lambda+1)}{(2\ell+1)!(2\lambda-2\ell +1)!}} \label{eq9} \\
&&  \hspace*{-5mm} (c_i)^{\lambda} (c_k)^{\lambda} (\sin \varphi_{ik})^{\lambda-\ell} (\cos \varphi_{ik})^\ell
\sum_{m_1 m_2}
\left ( \begin{array}{ccc} \ell&\lambda-\ell&\lambda \\ m_1 & m_2 &-\mu \end{array}\right)
\nonumber \\ &&  \hspace*{-5mm}
\langle n_0J_0M_0 | x_k^{\lambda-\ell} y_k^{\lambda+\ell}
 Y_{\lambda, \mu}^{*}(\Omega_{y_k})
Y_{\ell, m_1}(\Omega_{y_k})
 Y_{\lambda -\ell, m_2}(\Omega_{x_k}) |n_0J_0M_0 \rangle \;, \nonumber
\end{eqnarray}
\end{small}
where the constants 
\begin{equation}
c_i=\sqrt{\frac{m_N}{m_i} \frac{(m_j+m_k)}{M}} 
\label{ci}
\end{equation}
arise when inserting (\ref{eq5}) into the definition (\ref{eq1}).

The summation over the indexes $\mu$ and $m_1$ can be made analytically
\cite{rot59}, leading to  the final expression for the non-energy weighted sum rule:
\begin{eqnarray}
\label{eq10}
&& \hspace{-5mm} S_0 = \sum_{i,k=1}^3 \sum_{\ell=0}^{\lambda} z_i z_k
\left ( \begin{array}{ccc} \lambda&\lambda-\ell&\ell \\ 0 & 0 &0 \end{array}\right)
\frac{2\lambda+1}{\sqrt{2(\lambda-\ell)+1}} \\ &&\hspace{-5mm}
(c_i)^{\lambda} (c_k)^{\lambda}
 \sqrt{\frac{(2\lambda+1)!}{(2\ell)!(2\lambda-2\ell +1)!}} (\sin \varphi_{ik})^{\lambda-\ell} 
(\cos \varphi_{ik})^\ell \nonumber \\
&& \hspace{-5mm} \sum_{m_2}\langle n_0J_0M_0 | x_k^{\lambda-\ell} y_k^{\lambda+\ell}
Y_{\lambda-\ell, m_2}(\Omega_{x_k}) Y_{\lambda-\ell, m_2}^{*}(\Omega_{y_k}) |n_0J_0M_0 \rangle \; . \nonumber
\end{eqnarray}
The sum $(S_0)_{diag}$ of the diagonal terms ($i$=$k$) in
Eq.(\ref{eq10}) is obtained taking $\varphi_{ii}$=$\pi$ and
$\ell$=$\lambda$ (which reduce Eqs.(\ref{eq6}) and (\ref{eq8}) to
identities) which leads to:
\begin{equation}
 \left(S_0\right)_{diag} =  \frac{2\lambda+1}{4\pi} \sum_{i=1}^3 z_i^2
 \langle J_0 M_0| |\bm{r}_i -\bm{R}|^{2\lambda} |J_0M_0 \rangle \; .
\label{eq11}
\end{equation}

\begin{table}
\begin{center}
\caption{Non-energy weighted sum rule ($S_0$) and the contribution of 
the kinetic energy operator to the energy weighted sum rule
($S_1^{(T)}$) for a system of three particles with equal mass ($m$)
for $\lambda$=1. The first column gives the number $N$ of charged
particles each with charge $z$. The symbol $\langle \rangle$ denotes
expectation value in the initial state, and $\bm{r}_p$ is the
coordinate for one of the charged particles.}
\vspace*{0.5cm}
\begin{tabular}{|c|c|c|}
\hline
$N$ & $S_0$ & $S_1^{(T)}$ \\
\hline
$1$ & $3 z^2 \langle |\bm{r}_p -\bm{R}|^2 \rangle/(4 \pi)$ & $3 \hbar^2 z^2/(4 \pi m)$ \\
$2$ & $3 z^2 \langle |\bm{r}_p -\bm{R}|^2 \rangle/(4 \pi)$ & $3 \hbar^2 z^2/(4 \pi m)$ \\
$3$ & 0 & 0 \\
\hline
\end{tabular}
\label{tab1}
\end{center}
\end{table}

For a three-body system containing only one charged particle the
non-energy weighted sum rule reduces to one of the three diagonal
terms in Eq.(\ref{eq11}).  When more than one charged particle enter
in the three-body system, the full expression (\ref{eq10}), that
contains interferences between charged particles, must be used.

The relevance of the non-diagonal terms can be easily seen for a
system containing three identical particles with mass $m$ and charge
$z$ for $\lambda$=1. In this case the sum of the diagonal contributions $3
\frac{3 z^2}{4 \pi} \langle |\bm{r}_p -\bm{R}|^2 \rangle$ given by
Eq.(\ref{eq11}) is fully canceled by the non-diagonal terms, such that
$S_0$=0, as expected for three identical particles. 

 When only two of the particles
with mass $m$ each have the charge $z$, one of the diagonal terms is
canceled out by the non-diagonal one, and we get $S_0=\frac{3 z^2}{4
\pi} \langle |\bm{r}_p -\bm{R}|^2 \rangle$, which is identical to the
result when only one particle is charged.  These results are
summarized in the second column in table~\ref{tab1}.

\section{The energy-weighted sum rule}
\label{sec4}

The energy weighted sum rule is most easily obtained by evaluating the
expectation value of the double commutator in the ground state, i.e.
\begin{eqnarray}
&& S_1 =\frac{1}{2} \sum_{\mu}
 \langle n_0J_0M_0 |[[O_{\mu}^{\lambda\dagger},H],O_{\mu}^{\lambda}]
 |n_0J_0M_0 \rangle  = \nonumber \\ && \sum_{\mu}
 \langle n_0J_0M_0 |O_{\mu}^{\lambda\dagger} H O_{\mu}^{\lambda} -
 E_{0} O_{\mu}^{\lambda \dagger} O_{\mu}^{\lambda} |n_0J_0M_0 \rangle  \;,
\label{eq12}
\end{eqnarray}
where $E_{0}$ is the ground state energy.

This expression is obtained by inserting the identity operator
$\mathbbm{1}=\sum_{nJM}|nJM\rangle\langle nJM|$ between $H$ and
$O^\lambda_\mu$ and between $O^{\lambda\dagger}_\mu$ and
$O^\lambda_\mu$ , where $\{|nJM\rangle\}$ are the complete set of
eigenstates of $H$ with the corresponding set of eigenvalues
$\{E_{nJ}\}$.  In this way we immediately recover the standard
definition in Eq.(\ref{def}):
\begin{eqnarray}
  S_1 = \sum_{nJ} (E_{nJ} - E_{0}) \sum_{\mu M}
 |\langle n J M  | O_{\mu}^{\lambda}| n_0 J_0 M_0 \rangle|^2 \; .
\label{eq13}
\end{eqnarray}

According to Eq.(\ref{eq12}) the energy-weighted sum rule depends on
the multipole operator, the initial state properties, and the
hamiltonian. This hamiltonian can have a complicated angular momentum
dependence of both two- and three-body interactions, whose
contributions to $S_1$ in general do not vanish. In particular we
shall assume two-body interactions that depend on the relative partial
wave between the two particles, and three-body potentials
depending on the total angular momentum and parity of the three-body
state.

In the following subsections we evaluate the expression (\ref{eq12})
separately for the different terms of the hamiltonian, i.e. the
traditional contribution from the kinetic energy operator, and the
new terms arising from the partial wave dependent two-body potentials, and
$J^\pi$-dependent three-body potentials.

\subsection{Kinetic energy operator}

The kinetic energy operator can be expressed in terms of any of the
three sets of Jacobi coordinates as
\begin{equation}
T=-(\Delta_{x_k}+\Delta_{y_k})\hbar^2/2m_N \;,
\end{equation}
where the two Laplace operators $\Delta_{x_k}$ and $\Delta_{y_k}$ are
associated to the Jacobi coordinates $\bm{x}_k$ and $\bm{y}_k$. Since
the multipole operator (\ref{eq1}) only depends on
$\bm{y}$-coordinates we can quickly find the commutator 
$ [O_{\mu}^{\lambda\dagger},T]$ (Eq.(\ref{app1})), from which we get:
\begin{eqnarray}  \label{eq15} 
& [[O_{\mu}^{\lambda\dagger},T],O_{\mu}^{\lambda}] = \sum_{i,k=1}^3 z_i z_k
 \frac{\hbar^2}{m_N}   \\  \nonumber   
& \bm{\nabla}_{y_k} \left( |\bm{r}_k - \bm{R}|^{\lambda}
Y_{\lambda, \mu}^{*}(\Omega_{y_k})\right) \cdot 
  \bm{\nabla}_{y_k}
\left( |\bm{r}_i - \bm{R}|^{\lambda} Y_{\lambda, \mu}(\Omega_{y_i})\right) 
 \; .
\end{eqnarray}
Eq.(\ref{eq8}) permits rewriting of Eq.(\ref{eq15})
in terms of a single set of Jacobi coordinates, leading to:
\begin{eqnarray}
   &&   \hspace*{-5mm} [[O_{\mu}^{\lambda\dagger},T],O_{\mu}^{\lambda}] =  \frac{\hbar^2}{m_N} \sum_{i,k=1}^3
(c_i)^{\lambda} (c_k)^{\lambda} z_i z_k \sum_{\ell=0}^{\lambda}
\sum_{mn} (-1)^{-\mu} \nonumber \\ && \nonumber
 \times \sqrt{2\lambda+1} 
\left ( \begin{array}{ccc} \ell&\lambda-\ell&\lambda \\ m&n&-\mu
\end{array}\right) \sqrt{\frac{4\pi (2\lambda+1)!}{(2\ell+1)! 
 (2\lambda-2\ell+1)!}} \\ && \nonumber
 \times x_k^{\lambda-\ell} Y_{\lambda-\ell, n}(\Omega_{x_k}) 
(\sin \varphi_{ik})^{\lambda-\ell} (\cos \varphi_{ik})^\ell \\ &&
 \times \bm{\nabla}_{y_k} \left( y_k^{\lambda} 
 Y_{\lambda, \mu}^{*}(\Omega_{y_k})\right) \cdot
\bm{\nabla}_{y_k}
\left( y_k^{\ell} Y_{\ell, m}(\Omega_{y_k})\right) \label{eq17} \; .
\end{eqnarray}
The scalar product can now be performed by use of the Gradient Formula
(\ref{ap2}), and after writing the two spherical harmonics in terms
of a single one, and performing analytically the summations over angular
momentum projection quantum numbers (details are given in the appendix),
one gets the following final expression for
the contribution of the kinetic energy operator to the energy-weighted
sum rule $S_1^{(T)}$:
\begin{small}
\begin{eqnarray}
 S_1^{(T)} = \frac{-\hbar^2}{2m_N} \sum_{i,k=1}^3 
(c_i)^{\lambda} (c_k)^{\lambda} z_i z_k \sum_{\ell=1}^{\lambda}
 (\sin \varphi_{ik})^{\lambda-\ell} (\cos \varphi_{ik})^\ell
\label{eq19} \\ 
 \sqrt{\frac{\lambda \ell (2\ell-1) (2\lambda+1)!(2\lambda+1)^3}
        {(2\ell)! (2\lambda-2\ell+1) (2\lambda-2\ell+1)!}}
 \left ( \begin{array}{ccc} \lambda-1&\ell-1&\lambda-\ell \\ 0&0&0 \end{array}\right)
\nonumber \\ 
\sum_n \langle n_0J_0M_0 | x_k^{\lambda-\ell} y_k^{\lambda+\ell-2} Y_{\lambda-\ell, n}(\Omega_{x_k})
Y_{\lambda-\ell, n}^{*}(\Omega_{y_k}) |n_0J_0M_0 \rangle\;,  \nonumber
\end{eqnarray}
\end{small}
where the sum of the diagonal parts ($i$=$k$) become
\begin{eqnarray}
  \lefteqn{\left(S_1^{(T)}\right)_{diag}=
\frac{\hbar^2}{2m_N} \frac{\lambda (2\lambda+1)^2}{4 \pi} }
\label{eq20}  \\  && \times \sum_{i=1}^3 (c_i)^2 z_i^2 \langle n_0J_0M_0 |
|\bm{r}_i -\bm{R}|^{2\lambda-2} |n_0J_0M_0 \rangle \nonumber \; ,
\end{eqnarray}
where the constants $c_i$ are given by Eq.(\ref{ci}).

For $\lambda$=1 the expression in Eq.(\ref{eq19}) is independent of
the properties of the initial state. In particular, for three
identical particles with mass $m$ and charge $z$ the total value of
$S_1^{(T)}$ is zero, which confirms the result anticipated
at the end of section~\ref{sec2}.  When one of these
three particles has no charge, $S_1^{(T)}$ takes a constant value
which is the same as the one obtained when only one of the three
particles with mass $m$ is charged. The precise expressions of
$S_1^{(T)}$ for these particular cases are given in the last column of
table~\ref{tab1}.

\subsection{Partial-wave dependent two-body potentials}
\label{sub4b}

Typically, the two-body interactions are adjusted separately for the
individual partial waves in order to reproduce the known experimental
data for the two-body systems.  This procedure leads to two-body
interactions depending on the two-body quantum numbers $\{\ell_x,
s_x, j_x \}$.  The full two-body potential operator takes the form
$\hat{V}_{2b}=\sum_{i=1}^3 \hat{V}_{2b}^{(i)}$ where the index $i$
runs over all the three sets of Jacobi coordinates and
$\hat{V}_{2b}^{(i)}$ is the two-body operator describing the
interaction between particles $j$ and $k$.  This two-body operator is
formally written as:
\begin{eqnarray}
\hat{V}_{2b}^{(i)} & = & \sum_{\ell_{x_i},s_{x_i} }
 \sum_{j_{x_i},m_{x_i}}  V_{i}^{(\ell_{x_i},s_{x_i},j_{x_i})}(x_i) \hat{P}_i
\nonumber \\ & &
|\ell_{x_i},s_{x_i},j_{x_i},m_{x_i}\rangle \langle \ell_{x_i},s_{x_i},j_{x_i},m_{x_i}| \;,
\label{eq21}
\end{eqnarray}
where $\hat{P}_i$ represents any spin operator that could enter in the two-body potentials.

The contribution of the full two-body potential operator
$\hat{V}_{2b}$ to the second sum rule has then three contributions,
each corresponding to one of the three two-body
interactions. According to Eq.(\ref{eq12}), the contribution
$S_1^{(2b,i)}$ ($i$=1,2,3) from each of them is given by:
\begin{eqnarray}
S_1^{(2b,i)}&=&\sum_\mu \left[\langle n_0 J_0 M_0|O_\mu^{\lambda \dagger}
\hat{V}_{2b}^{(i)} O_\mu^\lambda|n_0 J_0 M_0 \rangle \right.
  \nonumber \\ & & \left.
 - \langle n_0 J_0 M_0|O_\mu^{\lambda \dagger} O_\mu^\lambda 
 \hat{V}_{2b}^{(i)}
|n_0 J_0 M_0 \rangle \right],
\label{eq26}
\end{eqnarray}
where $|n_0 J_0 M_0 \rangle$ represents the initial state with total
angular momentum $J_0$ and projection $M_0$. The quantum number $n_0$
refers to all other additional quantum numbers necessary to specify
this state.

For each two-body interaction $\hat{V}_{2b}^{(i)}$ it is now
convenient to write the corresponding ground state wave function
$\Psi$ in terms of the Jacobi coordinates $\{\bm{x}_i,\bm{y}_i\}
\equiv \{\rho,\alpha_i,\Omega_{x_i},\Omega_{y_i}\}$, and expand it in
terms of a set of
functions ${\cal Y}_{\gamma_i}^{J_0M_0}(\Omega_i)$
\begin{equation}
\Psi^{J_0M_0}_{n_0}(\bm{x}_i,\bm{y}_i)=\frac{1}{\rho^{5/2}}
\sum_{\gamma_i} F_{\gamma_i}^{n_0J_0}(\rho) {\cal Y}_{\gamma_i}^{J_0M_0}(\Omega_i),
\label{eq27}
\end{equation}
where
\begin{equation}
{\cal Y}_{\gamma_i}^{J_0M_0}(\Omega_i)=\phi_K^{(\ell_{x_i},\ell_{y_i})}(\alpha_i)
\left[  
|\ell_{x_i},s_{x_i},j_{x_i} \rangle\ \otimes  |\ell_{y_i},s_i,j_{y_i} \rangle\
 \right]^{J_0M_0}
\label{eq28}
\end{equation}
with $\gamma_i \equiv \{K,\ell_{x_i},s_{x_i},j_{x_i},\ell_{y_i},j_{y_i} \}$, and
with $\phi_K^{(\ell_{x_i},\ell_{y_i})}(\alpha_i)$ being the usual function
of the hyperangle $\alpha_i$ entering in the definition of the hyperspherical
harmonics \cite{nie01}. The functions (\ref{eq28}) reduce to the usual hyperspherical
harmonics for particles without spin.

With the definition (\ref{eq21}), the two-body potential operator $\hat{V}_{2b}^{(i)}$
acting on a term of the basis ${\cal Y}_{\gamma_i}^{J_0M_0}(\Omega_i)$ (written in
the Jacobi set $i$) leads to
\begin{equation}
\hat{V}_{2b}^{(i)}{\cal Y}_{\gamma_i}^{J_0M_0}(\Omega_i)=
  V_{i}^{(\ell_{x_i},s_{x_i},j_{x_i})}(x_i) \hat{P}_i
  {\cal Y}_{\gamma_i}^{J_0M_0}(\Omega_i)
\label{eq29}
\end{equation}

Eq (\ref{eq27}) permits to write Eq.(\ref{eq26}) as:
\begin{eqnarray}
S_1^{(2b,i)} &\!\!\!\!\!=\!\!\!\!\!&
\sum_\mu \int d\rho \sum_{\gamma_i} \sum_{\gamma_i^\prime}
F_{\gamma_i}^{n_0 J_0}(\rho) F_{\gamma_i^\prime}^{n_0 J_0}(\rho)
  \label{eq30} \\ & &  \hspace*{-1.2cm}
\left[\langle {\cal Y}_{\gamma_i}^{J_0 M_0}| O_\mu^{\lambda \dagger} \hat{V}_{2b}^{(i)} 
O_\mu^\lambda|{\cal Y}_{\gamma_i^\prime}^{J_0 M_0}\rangle
-\langle {\cal Y}_{\gamma_i}^{J_0 M_0}|O_\mu^{\lambda \dagger} O_\mu^\lambda \hat{V}_{2b}^{(i)}|
{\cal Y}_{\gamma_i^\prime}^{J_0 M_0}\rangle\right]  \nonumber
\end{eqnarray}

Inserting in the first and second matrix elements the unity operator
\begin{equation}
\mathbbm{1}=\sum_{\gamma_i''} \sum_{J''M''} 
| {\cal Y}_{\gamma_i''}^{J'' M''}(\Omega_i)   \rangle 
\langle {\cal Y}_{\gamma_i''}^{J'' M''}(\Omega_i)|,
\end{equation}
between $O_\mu^{\lambda \dagger}$ and $\hat{V}_{2b}^{(i)}$, and
$O_\mu^{\lambda \dagger}$ and $O_\mu^{\lambda}$, respectively, and making use
of Eq.(\ref{eq29}), we immediately get the final expression:
\begin{eqnarray}
S_1^{(2b,i)}&=&\sum_{\mu} \int \ud \rho \sum_{\gamma_i} \sum_{\gamma_i^\prime}
F_{\gamma_i}^{n_0J_0}(\rho) F_{\gamma_i^\prime}^{n_0J_0}(\rho) \nonumber \\
&& \hspace*{-1cm}
\sum_{\gamma_i''}\sum_{J'' M''} \langle {\cal Y}_{\gamma_i}^{J_0 M_0}(\Omega_i)|
O_\mu^{\lambda \dagger}|{\cal Y}_{\gamma_i''}^{J'' M''}(\Omega_i)\rangle
\nonumber \\ && \hspace*{-1cm}
\langle {\cal Y}_{\gamma_i''}^{J'' M''}(\Omega_i)|
\left(V_{i}^{(\ell''_{x_i},s''_{x_i},j''_{x_i})}(x_i) \hat{P}_i O_\mu^\lambda 
\right.
\nonumber \\ && \hspace*{-1cm}
\left.
-V_{i}^{(\ell'_{x_i},s'_{x_i},j'_{x_i})}(x_i) 
  O_\mu^\lambda \hat{P}_i \right) |{\cal Y}_{\gamma_i^\prime}^{J_0 M_0}(\Omega_i)\rangle \;. \label{eq32}
\end{eqnarray}

When the $V_i$-functions are independent of the partial wave, the equation above
can be written in a more compact way as:
\begin{eqnarray}
S_1^{(2b,i)} &\!\!\!\!\!=\!\!\!\!\!& \frac{1}{2}
\sum_\mu \int d\rho \sum_{\gamma_i} \sum_{\gamma_i^\prime}
F_{\gamma_i}^{n_0J_0}(\rho) F_{\gamma_i^\prime}^{n_0J_0}(\rho)
 \\ & &  \hspace*{-1.2cm}
\langle {\cal Y}_{\gamma_i}^{J_0 M_0}| 
V_i(x_i)[[O_\mu^{\lambda \dagger},\hat{P}_i],O_\mu^\lambda]
 | {\cal Y}_{\gamma_i^\prime}^{J_0 M_0}\rangle  \nonumber
\end{eqnarray}
which is trivially zero for the central part of the two-body potential ($\hat{P}_i=1$), and
for the spin-spin term ($O_\mu^\lambda$ does not depend on the spin and therefore commutes
with the spin-spin operator). The same happens for the tensor operator, which depends only on
coordinates and spin operators.  For the spin-orbit term ($\bm{\ell}_x\cdot \bm{s}_x=
\ell_+s_-+\ell_-s_++\ell_zs_z$) one has the same result, since $\ell_+$, $\ell_-$, or $\ell_z$ applied on
$O^\lambda_\mu$ is proportional to either $O^\lambda_{\mu+1}$, $O^\lambda_{\mu-1}$,
or $O^\lambda_{\mu}$, and therefore each of the three terms in $\bm{\ell}_x\cdot \bm{s}_x$
double commutes with $O^\lambda_\mu$.

Thus, for two-body interactions independent of the partial waves
and containing the usual spin operators one has $S_1^{(2b)}=0$.

If the the two-body potentials are partial wave dependent, but contain only 
central, spin-spin ($\hat{P}_i$=$\bm{s}_j \cdot \bm{s}_k$) and spin-orbit 
$(\hat{P}_i$=$\bm{\ell}_{x_i} \cdot \bm{s}_{x_i})$ terms, since these
operators are diagonal in the basis $\{|\ell_{x_i},s_{x_i},j_{x_i},m_{x_i} \rangle\}$,
one then has:
\begin{equation}
\hat{P}_i{\cal Y}_{\gamma_i}^{J_0M_0}(\Omega_i)=
  f_{\ell_{x_i},s_{x_i}}^{j_{x_i}} {\cal Y}_{\gamma_i}^{J_0M_0}(\Omega_i)
\end{equation}
where $f_{\ell_{x_i},s_{x_i}}^{j_{x_i}}$=1 for the central part of the potential,
$f_{\ell_{x_i},s_{x_i}}^{j_{x_i}}=(s_{x_i}(s_{x_i}+1)-s_j(s_j+1)-s_k(s_k+1))/2$ for the
spin-spin part, and $f_{\ell_{x_i},s_{x_i}}^{j_{x_i}}=(j_{x_i}(j_{x_i}+1)
-\ell_{x_i}(\ell_{x_i}+1)-s_{x_i}(s_{x_i}+1))/2$ for the spin-orbit part. 
Eq.(\ref{eq32}) can then be written for this particular case as:
\begin{eqnarray}
S_1^{(2b,i)}&=&\sum_{\mu} \int \ud \rho \sum_{\gamma_i} \sum_{\gamma_i^\prime}
F_{\gamma_i}^{n_0J_0}(\rho) F_{\gamma_i^\prime}^{n_0J_0}(\rho) \nonumber \\
&& \hspace*{-1cm}
\sum_{\gamma_i''}\sum_{J'' M''} \langle {\cal Y}_{\gamma_i}^{J_0 M_0}(\Omega_i)|
O_\mu^{\lambda \dagger}|{\cal Y}_{\gamma_i''}^{J'' M''}(\Omega_i)\rangle
\nonumber \\ && \hspace*{-1cm}
\langle {\cal Y}_{\gamma_i''}^{J'' M''}(\Omega_i)|
\left(V_{i}^{(\ell''_{x_i},s''_{x_i},j''_{x_i})}(x_i) f_{\ell''_{x_i},s''_{x_i}}^{j''_{x_i}}
\right.
\nonumber \\ && \hspace*{-1cm}
\left.
-V_{i}^{(\ell'_{x_i},s'_{x_i},j'_{x_i})}(x_i) f_{\ell'_{x_i},s'_{x_i}}^{j'_{x_i}} \right)
  O_\mu^\lambda |{\cal Y}_{\gamma_i^\prime}^{J_0 M_0}(\Omega_i)\rangle \label{eq30b}\;. 
\end{eqnarray}

It is important to keep in mind that the operator $O^\lambda_\mu$ has
three terms (see Eq.(\ref{eq1})), each expressed in one of the three
sets of Jacobi coordinates. When inserted in Eq.(\ref{eq30b}), the two
terms in $O^\lambda_\mu$ differing from the set of Jacobi coordinates
$i$ must be transformed into this set by use of Eq.(\ref{eq8}).

When only one of the three particles is charged the operator
$O^\lambda_\mu$ reduces to one term. A partial wave dependence in the
interaction between the charged particle and any of the other two will
produce a non-vanishing contribution to the energy-weighted sum rule
according to Eq.(\ref{eq30b}). However, if the only partial wave dependence
appears in the two-body potential between the two neutral particles then
$S_1^{(2b)}$=0. This is because the $O^\lambda_\mu$ operator then
automatically is written in the same Jacobi set as the angular
functions in Eq.(\ref{eq30b}).  The operator is then independent of
$\Omega_{x_i}$, and the integral over these angles in the last matrix
element of (\ref{eq30b}) vanishes unless $\ell''_{x_i}=\ell'_{x_i}$,
$s''_{x_i}=s'_{x_i}$, and $j''_{x_i}=j'_{x_i}$
and therefore the full matrix element vanishes.

The integrals over $\Omega_{x_i}$ and $\Omega_{y_i}$ in the two matrix
elements that appear in Eq.(\ref{eq30b}) can be calculated analytically,
because $x_i=\rho \sin \alpha_i$, and therefore the two-body
potentials are independent of the angles $\Omega_{x_i}$ and
$\Omega_{y_i}$.  The expressions for these two matrix elements are
given as Eq.(\ref{ap6}) of the appendix for the particular case
of particles without spin.

\begin{figure}
\epsfig{file=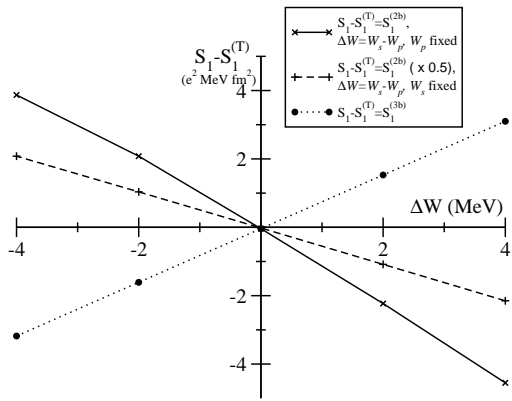, width=6.5cm, angle=-90}
\caption{For dipole excitations in a system of three particles 
with equal mass ($m$=4 times the nucleon mass), and only one of them
with a charge (equal to twice the proton charge), the figure shows the
variation of $(S_1-S_1^{(T)})$ as a function of the strength
difference ($\Delta W$) between the gaussian $s$- and $p$-wave
two-body potentials of equal range (see text). The solid and dashed 
lines give the results when only the
$s$-wave strength and only the $p$-wave strength are changed,
respectively. The dotted line shows the same variation as a function
of the strength difference between the gaussian three-body forces in
the excited (1$^-$) and ground ($0^+$) states when the three-body
potentials have equal range, when only the strength of the 1$^-$
states is changed, and when the two-body potentials are
$\ell$-independent.}
\label{fig1}
\end{figure}

As an example we consider dipole excitations ($\lambda$=1) in a system
of three spin zero particles with equal mass $m$=4$m_N$, where $m_N$ is the
nucleon mass, and where two particles are neutral and one particle has
a charge equal to twice the proton charge.  We consider only $s$ and
$p$ waves in the calculation. The two-body interactions are taken to
be gaussians ($V_{s,p}(r)=W_{s,p}e^{-r^2/b^2}$) with equal range $b$
for $s$ and $p$ waves. We have
constructed a $0^+$ ground state with a very large contribution of
$s$-waves, and a few percent of $p$-waves.  This has been done by taking
$b$=2.98 fm, $W_s$=$W_p$=$-0.18$ MeV for the interaction between the two
neutral particles, and $W_s$=$W_p$=$-1.18$ MeV for the interaction between
the charged particle and one of the neutral ones. The binding energy of the
$0^+$ state is $-10.38$ MeV. According to Eq.(\ref{eq30b}), since only terms with
$\ell''_{x_i}\neq\ell'_{x_i}$ contribute ($s''_{x_i}$=$s'_{x_i}$=0,
$j''_{x_i}$=$\ell''_{x_i}$, $j'_{x_i}$$=\ell'_{x_i}$, and
$f_{\ell_{x_i},s_{x_i}}^{j_{x_i}}$=1), the contribution $S_1^{(2b)}$ to the 
energy-weighted sum rule is proportional to $\Delta W=W_s-W_p$, where
$W_s$ and $W_p$ refer to the strengths of the interactions between the
charged and the neutral particle.

In Fig.\ref{fig1}, the dashed line ($+$ signs) shows $S_1^{(2b)}$ as a
function of $\Delta W$ when the strength of the $s$-wave potential
between the charged and neutral particles
($W_s$) is kept fixed and $W_p$ is changed. Since the $p$-wave
contribution to the ground state wave function is insignificant, a
small variation in the strength of the $p$-wave potential only
slightly modifies the ground state radial wave functions
$F_{\gamma}^{n_0J_0}(\rho)$. Therefore, the behaviour of
$S_1^{(2b)}$ is almost perfectly linear with $\Delta W$. However, if
we modify $\Delta W$ by keeping $W_p$ fixed while changing $W_s$, the
radial wave functions are much more sensitive to a change in the
$s$-wave two-body potential, since the $s$-waves dominate.  Therefore
$S_1^{(2b)}$ is not a completely linear function of $\Delta W$ as seen
by the solid line (--$\times$ signs-- in the figure).  When $\Delta
W$=0, the total value of $S_1$ for this particular case is $S_1$=9.90
$e^2$ MeV fm$^2$, which means that the contribution from $\Delta
W\neq$0 can be of comparable size, see Fig.\ref{fig1}.

\subsection{$J^\pi$-dependent three-body potentials}

When performing three-body calculations it is quite usual to employ
effective three-body forces to fine tune the energies of the computed
states. Very often different three-body forces are used to place the
lowest state with given angular momentum and parity $J^\pi$ at the
correct energy.  This means that these three-body potentials usually
depend on $J^\pi$. In this subsection we investigate the additional
contribution $S_1^{(3b)}$ to the energy-weighted sum rule arising from
this kind of three-body potentials.  With these assumptions the
three-body potential operator can be written as:
\begin{equation}
\hat{V}_{3b} = \sum_{J \pi M} \sum_{n_{(J^\pi)}}  V^{(J^\pi)}_{3b}(\rho)  |n_{(J^\pi)}J^\pi M \rangle
    \langle n_{(J^\pi)}J^\pi M|
\label{eq35}
\end{equation}
where $n_{(J^\pi)}$ refers to all the additional quantum numbers
needed to specify each of the three-body states with total angular
momentum and parity $J^\pi$.  Following Eq.(\ref{eq12}) we can write:
\begin{eqnarray}
 S_1^{(3b)} & = &
 \sum_{\mu} \left[
 \langle n_0J_0^{\pi_0}M_0 |O_{\mu}^{\lambda\dagger} \hat{V}_{3b} O_{\mu}^{\lambda}|n_0J_0^{\pi_0}M_0 \rangle
\right. \nonumber \\ & - &
 \left. \langle n_0J_0^{\pi_0}M_0 |O_{\mu}^{\lambda \dagger}
          O_{\mu}^{\lambda} \hat{V}_{3b}|n_0J_0^{\pi_0}M_0 \rangle \right] \;,
\label{eq36}
\end{eqnarray}
where we explicitly labeled the initial state by its parity $\pi_0$.

Substituting now Eq.(\ref{eq35}) into (\ref{eq36}) and inserting
the unity operator between $O_{\mu}^{\lambda \dagger}$ and
$O_{\mu}^{\lambda}$ in the last matrix element, we finally get:
\begin{eqnarray} \label{eq31}
S_1^{(3b)}&=&\sum_{\mu} \sum_{J \pi M}\sum_{n_{(J^\pi)}}
\langle n_{(J^\pi)}J^\pi M| O_{\mu}^{\lambda} |n_0 J_0^{\pi_0} M_0 \rangle  
  \\  \nonumber 
 &\times& \langle n_0 J_0^{\pi_0}M_0| O_{\mu}^{\lambda \dagger} 
(V^{(J^\pi)}_{3b}-V^{(J_0^{\pi_0})}_{3b})
|n_{(J^\pi)}J^\pi M \rangle  \;\;\;
\end{eqnarray}
that gives the contribution to the energy-weighted sum rule from
$J^\pi$-dependent three-body potentials. This contribution vanishes
when the three-body interactions are $J^\pi$-independent.

For the special case in which the ground state has $J_0=0$ the
expression simplifies to:
\begin{small}
\begin{equation}
S_1^{(3b)} =
\sum_{\mu} \langle n_0 0^{ \pi_0} 0 | |O_{\mu}^{\lambda}|^2
 \left( V^{(\lambda^{\pi})}_{3b}(\rho) - V^{(0^{\pi_0})}_{3b}(\rho) \right)
 | n_0 0^{\pi_0} 0 \rangle   \;,
\label{eq39}
\end{equation}
\end{small}
where $\pi=\pi_0(-1)^\lambda$ and which, except for the difference 
between the three-body potentials, is
similar to Eq.(\ref{eq3}).  Therefore, the analytic expression of
$S_1^{(3b)}$ for $J_0$=0 is given by Eq.(\ref{eq10}), but with an
additional factor equal to the difference between the three-body
potentials inserted in the last matrix element.

In Fig.\ref{fig1} the dotted line (with circles) shows $S_1^{(3b)}$
for the same system and the same transition as for the $S_1^{(2b)}$
case. We have taken the $\ell$-independent two-body potentials used in 
subsection \ref{sub4b} as starting point, meaning
that $S_1^{(3b)}=S_1-S_1^{(T)}$. The result is shown as a function of
the strength difference ($\Delta W$) between the gaussian effective
three-body forces used to compute the $1^-$ excited states and the $0^+$
ground state. The range of the three-body force is the same (6.0 fm) for
$0^+$ and $1^-$.  The variation in $\Delta W$ is obtained by changing
the strength in the three-body force for the $1^-$ excited states.
Then the ground state wave function remains unchanged. As a
consequence, according to Eq.(\ref{eq31}), and as demonstrated by the
dotted line in the figure, $S_1^{(3b)}$ depends linearly on $\Delta
W$. The contribution from $S_1^{(3b)}$ can be of comparable size to
the value, $S_1$=9.90 $e^2$ MeV fm$^2$, for angular momentum
independent potentials.

\section{A realistic case: Dipole excitations in $^6$He}
\label{sec5}

The main properties of the borromean two-neutron halo nuclei are well
reproduced describing them as three-body systems made by an inert core
surrounded by two neutrons. The characteristic feature of these nuclei
is their large spatial extension, which is responsible for the large
values of the breakup cross sections after electromagnetic excitation.
This can be easily envisaged from Eqs.(\ref{eq11}) and (\ref{eq20}),
which depend directly on the size of the system.  For this reason,
electromagnetic excitations of two-neutron halo nuclei have attracted
a lot of attention, specially dipole excitations, which is the
dominating multipolarity for such excitations.

In this section we investigate dipole excitations in $^6$He
($\alpha$+$n$+$n$), which is one of the most prominent examples of
borromean two-neutron halo nuclei. We compute the three-body states by
use of the hyperspheric adiabatic expansion method \cite{nie01}. The
neutron-neutron and $\alpha$-neutron interactions are the ones used
for instance in \cite{gar99}.  The computed bound ground state ($0^+$)
has a two-neutron separation energy matching the experimental value of
$-0.97\pm 0.04$ MeV. This is achieved with a gaussian effective
three-body force with range 2.9 fm and strength $-7.55$ MeV. The
continuum 1$^-$ states have been discretized by use of a box boundary
condition at $\rho_{max}$=50 fm.

\subsection{Dipole strength function}

The transition probability $\cal B$ from the ground state to one of the
box discretized continuum states is given in Eq.(\ref{eq77}). To obtain
a smooth distribution from the discretized continuum we use the finite
energy interval approximation to the strength function,
\begin{equation}
\frac{d{\cal B}}{dE} \approx \frac{\Delta {\cal B}}{\Delta E} \;,
\end{equation}
where $\Delta E$ is the size of the given energy interval, and
$\Delta{\cal B}$ is the sum of the transition probabilities into the
states whose energies fall into this interval. These values are then
plotted as function of the central energy values of the intervals.

The interval should be of a reasonable size, i.e. large enough to provide a 
smooth function but small enough not to wash out the desired
structure. In practice we have used bins with centers at 0.3, 0.9,
1.5, 2.3, 3.2, 4.2, 5.4, 6.7, and 8.2 MeV, and a standard
cubic interpolation to smooth the curve.

\begin{figure}
\epsfig{file=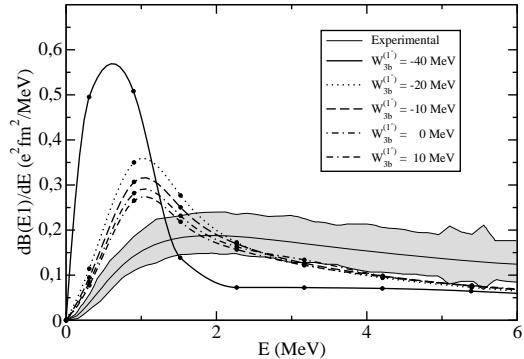, width=6cm, angle=-90}
\caption{Computed (circles) dipole strength function for $^6$He. The curve
through the computed points is obtained by cubic interpolation. Gaussian
three-body forces are used to obtain the ground ($0^+$) and excited
($1^-$) three-body states. The range of the gaussian (2.9 fm) is the same for
both angular momenta.  The different curves show the results obtained
with different values of the strength ($W_{3b}^{(1^-)}$) of the
three-body force for the 1$^-$ states. The experimental data 
(shaded area) are from \cite{aum99}. }
\label{fig2}
\end{figure}

In Fig.~\ref{fig2} we compare with the measured distribution \cite{aum99} 
the computed smoothed dipole strength function for $^6$He for different gaussian
three-body potentials in the excited $1^-$ states. A strongly attractive three-body
potential produces a pronounced low-lying peak which for even stronger
attraction would turn into a bound state.  Thus the three-body
potential should at least be less attractive than that corresponding
to $W^{(1^-)}_{3b} = -40$~MeV ($b$=2.9 fm).  For moderately attractive three-body
potentials we observe an increase from zero at threshold to a peak
value at around $1$~MeV followed by a relatively fast decrease towards
zero at higher energies. This is consistent with the calculations in
\cite{myo01} and~\cite{dan98}.  Compared to the experiment \cite{aum99},
the theory overestimates the strength at around $1$~MeV and consistent
with the sum rule underestimate the strength at higher energies.
Apparently a significant three-body repulsion in the $1^-$ channel
would approach the experimental data presented with relatively large
error bars.

\subsection{Sum rule results}

\begin{table}
\caption{Non-energy weighted ($S_0$) and energy weighted ($S_1$) 
dipole sum rule values for $^6$He. The fourth column gives the
contribution to $S_1$ from the kinetic energy operator
($S_1^{(T)}$). The last column ($S_1/S_0$) is an average dipole
resonance energy.  The upper and lower part of the table show the
experimental and computed sum rule results for states below energies
of 5 MeV and 10 MeV, respectively.  The last row gives the converged
results including all excitations.  The experimental data are from
\cite{aum99}.  The $S_0$ values are given in units of $e^2$ fm$^2$ and
$S_1$ and $S_1^{(T)}$ are in units of $e^2$ fm$^2$ MeV. The average
dipole resonance energy $S_1/S_0$ is given in MeV.}
\vspace*{5mm}
\begin{tabular}{|c|cccc|}
   \hline
                          &  $S_0$         &  $S_1$  &  $S_1^{(T)}$ & $S_1/S_0$ \\ \hline
$E^*\leq 5$ MeV (exper.)  & 0.59$\pm$0.12  & 1.9$\pm$0.4  & --   & 3.22$\pm$0.94 \\
$E^*\leq 5$ MeV (theor.)  & 0.66           & 1.94         & --   & 2.94 \\ \hline
$E^*\leq 10$ MeV (exper.) & 1.2$\pm$0.2    & 6.4$\pm$1.3  & --   & 5.3$\pm$1.4 \\
$E^*\leq 10$ MeV (theor.) & 1.01           & 4.43         & --   & 4.39 \\ \hline
Converged                 & 1.25           & 8.26         & 4.95 & 6.61 \\ \hline
\end{tabular}
\label{tab2}
\end{table}

The second and third columns of table~\ref{tab2} give the non-energy
weighted ($S_0$) and energy-weighted ($S_1$) dipole sum rule
strengths. The experimental data, available from \cite{aum99}, are
given in the first and third rows including states of energies below 5
MeV and 10 MeV, respectively. The corresponding theoretical values are
obtained numerically directly from the first row of Eq.(\ref{eq3}) and
Eq.(\ref{eq13}), and they are given in the second and fourth rows of
the table.

We can see that the computed results for $S_0$ agree very well with
the experimental values when the sum over the excited states in
Eq.(\ref{eq3}) is restricted to energies below 5 MeV and 10 MeV,
respectively. As even higher energies are included the value of $S_0$
converges to the result given in the last row of the table, which
agrees with the expected result obtained from Eq.(\ref{eq11}). The
converged value is already reached with an energy limit of about 40
MeV.

Essentially the same happens for $S_1$. The computed values agree
reasonably well with the experimental ones. Also the result obtained
for energies below 10 MeV is still clearly below the converged value,
which requires integration up to energies at least of about 60
MeV. The converged value for $S_1$ clearly disagrees with the result
provided by Eq.(\ref{eq20}), where only the kinetic energy
contribution is considered. This value is given in the fourth column
of table~\ref{tab2}. It is important to note that the computed results
given in the table have been obtained using the same effective
three-body force for the $0^+$ ground state and the $1^-$ excited
states. This means that the difference between the converged $S_1$
value and $S_1^{(T)}$ is exclusively due to the effect of the
$\ell$-dependence of the two-body $\alpha$-neutron potentials (see
Eq.(\ref{eq30b})). As seen in the table, this effect is far from being
negligible.

The last column in table~\ref{tab2} shows the ratio between the energy
and non-energy weighted sum rules, which is interpreted as an average
energy of the soft dipole mode. The value of 6.6 MeV obtained after
reaching convergence in $S_0$ and $S_1$ is consistent with previous
results, like \cite{suz91,dan93}, where a value of about 5 MeV also is
obtained. In \cite{dan98} a clearly smaller value is given (3.8 MeV),
very likely because they used $S_1^{(T)}$ instead of the full $S_1$ in
the computation of the ratio.  In any case these rather large
variations illustrate how important it is in practice to use the
correct sum rules in such estimates.

So far all the calculations of the $S_1$ strength have been performed
with the same effective three-body force for the ground state and the
$1^-$ excited states. This force was adjusted to fit the experimental
two-neutron separation energy in the $0^+$-state. However, for the
$1^-$-states the interaction might be different and an additional
contribution to the $S_1$ strength would appear as seen in
Eq.(\ref{eq39}).

\begin{figure}
\epsfig{file=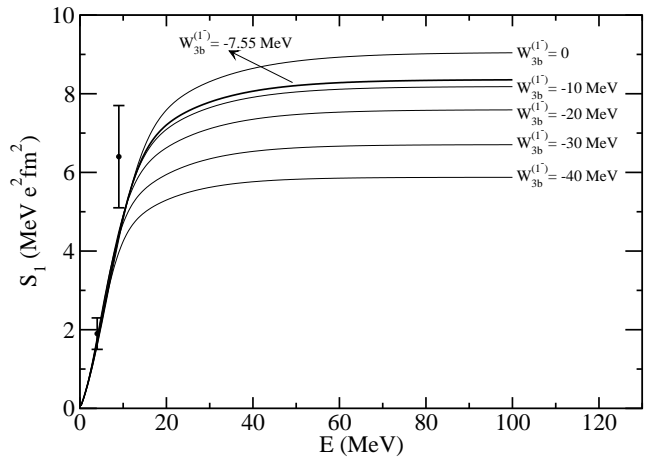, width=6cm, angle=-90}
\caption{Computed dipole sum rule strength $S_1$ for $^6$He as a 
function of the maximum energy allowed above threshold. Gaussian
three-body forces are used to obtain the ground ($0^+$) and excited
($1^-$) three-body states. The range of the gaussian is the same (2.9 fm) for
all the states. The different curves show the results obtained with
different values of the strength ($W_{3b}^{(1^-)}$) of the three-body
force for the 1$^-$ states. The thick curve is the result with the
same three-body force for the ground and excited states. The
experimental data are from \cite{aum99}.}
\label{fig3}
\end{figure}

In Fig.~\ref{fig3} we show the dipole sum rule strength $S_1$ for
$^6$He as a function of the maximum energy allowed above threshold.
We have considered gaussian three-body forces with a range of 2.9
fm. For the ground state (0$^+$) a strength of $-7.55$ MeV has been
used. When the excited states are obtained with the same three-body
force, $S_1$ behaves as shown by the thick solid line in the
figure. This calculation corresponds to the numbers quoted in
table~\ref{tab2}. When the strength of the three-body force used for
the $1^-$ states is changed, $S_1^{(3b)}$ depends linearly on the
strength difference.

The thin solid lines in the figure show $S_1$ for different values of
the strength ($W_{3b}^{(1^-)}$) of the three-body force for the
$1^-$-states. This strength has been changed from 0 up to $-40$ MeV,
which is at the limit of producing a low-lying narrow 1$^-$-resonance
in $^6$He (understood as a pole of the $S$-matrix). As seen in the
figure, the deeper the three-body potential, the smaller the value of
$S_1$.  The converged value can change significantly with the
three-body force. The result obtained with a strength of $-40$ MeV is
about 33\% smaller than obtained for $W_{3b}^{(1^-)}$=0.  Therefore,
the value of the average dipole resonance energy (the ratio between
the values of the two sum rules) also changes substantially with the
three-body force, ranging between 7.2 MeV when $W_{3b}^{(1^-)}=0$, and
4.7 MeV when $W_{3b}^{(1^-)}=-40$ MeV.

As seen in the figure, all the curves agree well with the experimental
value obtained for states below a maximum energy of 5 MeV. When this
maximum energy is 10 MeV, all the computed curves are below the
experiment. Although this experimental value has a rather large error
bar, it is clear from the figure that the smaller strengths in the
three-body force for the $1^-$ states are closer to the experimental
value.  This tendency is consistent with the fact that a
$1^-$-resonance has not been found experimentally since a strength
weaker than about $-30$~MeV also excludes such a state in
computations. Extension to include higher energies in the experiment
would allow distinction between the values obtained for different
strengths. Probably $^6$He is a very favorable system for this
investigation because the core excitations are expected to be
negligible.

\section{Summary and conclusions}
\label{sec6}

We have derived general expressions for the non-energy and
energy-weighted cluster sum rules for excitations of three-body
systems.  We consider transitions arising due to electric multipole
operators of any order and each of the constituent particles
(clusters) may have a finite charge.  The most obvious nuclear
applications are in systems close to three-body thresholds where
three-body clusterization frequently seems to be a dominating part of
the structure.  This is also the region where the spatially extended
and weakly bound halos appear.

Accurate calculations of three-body wave functions require in general
decomposition into Faddeev components either by directly solving the
Faddeev equations or by a variational procedure including similar
components expressed in the different Jacobi coordinates.  Derivation
of the non-energy weighted sum rule only relies on the use of a
complete set of intermediate wave functions.  Therefore only matrix
elements of the multipole operators enter into the expressions whereas
the interactions disappear altogether, except of course indirectly
through the properties of the excited continuum states. However, the
properties of the wave functions are essential and in particular the
different Faddeev components give rise to crucial interference effects
when more than one particle carry a charge.

Such interference effects are also crucial for the energy-weighted sum
rule where in addition also the properties of the interactions are
essential.  This sum rule is traditionally derived as a double
commutator between the hamiltonian and two multipole operators.
Usually then only the second order derivatives from the kinetic energy
operator contribute while the potentials including the spin-orbit
terms commute with the multipole operators and lead to vanishing
contributions.  However, when the interactions are angular momentum
dependent, the double commutator does not vanish because the multipole
operators themselves also carry angular momentum.  These contributions
must therefore be computed and included in the sum rule estimates.
Still the character of sum rule remains in the sense that no matter
how the excitations are distributed, they must add up to the value
given by the sum rule which only depends on properties of the
ground state and the interactions.

The angular momentum dependence and the subsequent contributions to
the sum rule are separated into terms arising from the two- and
three-body potentials which often in accurate three-body computations
depend on angular momentum. A possible sequence to determine
appropriate potentials could be first to adjust the two-body
potentials independently for each partial wave to known two-body bound
or continuum properties. Second to fine-tune the three-body state
computed with the two-body potentials to a desired energy by adding a
short-range three-body potential with as little structure as possible
in order to maintain the properties provided by the two-body
interactions.  Both types of angular momentum dependence are important
as they turn out to give substantial contributions to the ordinary
kinetic energy contribution to the sum rule.

To assess numerically the relative importance of these new sum rule
contributions we investigate the electric dipole excitations of the
ground state of the well known halo nucleus $^6$He.  We first notice
that the strength distribution has a peak at around 1 MeV, falls off
at higher energies and in practice reaches zero at about 60~MeV. The
contribution to the energy weighted sum rule from the two-body
potentials amounts to 2/3 of the kinetic energy contribution. 

The contribution from the three-body potential depends on an
expectation value of the difference between those potentials for
ground and excited state angular momenta of 0 and 1, respectively.
Thus for state independent but finite three-body potentials we arrive
at the established result of zero contribution.  However, the
sensitivity to the difference in these three-body potentials is
significant. Realistic potentials give estimates of up to 30\% of the
kinetic energy value.  This is then also an estimate of the
sensitivity of the soft dipole mode to the three-body potential.

In conclusion, we have generalized the energy weighted and non-energy
weighted cluster sum rules for electric multipole transitions to
angular momentum two- and three-body interactions. The additional
contributions can be comparable in size to the ordinary terms arising
from the kinetic energy operator. 

\section*{Acknowledgments}

This work was partly supported by funds provided by DGI of MEC (Spain)
under contract No. FIS2005-00640.
One of us (R.D.) acknowledges support by a predoctoral I3P fellowship
from CSIC and the European Social Fund.

\appendix
\section{Intermediate expressions and formulas}

In this appendix we give some of the intermediate expressions obtained when
deriving Eq.(\ref{eq19}).

The commutator between the electric multipole operator $O_{\mu}^{\lambda\dagger}$
and the kinetic energy operator $T$ can be written as:
\begin{eqnarray}
& & [O_{\mu}^{\lambda\dagger},T] = -\sum_{k=1}^3 z_k \frac{\hbar^2}{2 m_N}
[|\bm{r}_k - \bm{R}|^{\lambda} Y_{\lambda, \mu}^{*}
 (\Omega_{y_k}),\Delta_{y_k}] \; \nonumber \\
& & =\sum_{k=1}^3 z_k \frac{\hbar^2}{m_N} \bm{\nabla}_{y_k}
 \left( |\bm{r}_k - \bm{R}|^{\lambda} Y_{\lambda, \mu}^{*}(\Omega_{y_k})\right)\cdot
 \bm{\nabla}_{y_k} \; ,
\label{app1}
\end{eqnarray}
where we have used that $\Delta_k(r_k^j Y_{j,m}(\Omega_k))$=0.

The scalar product in Eq.(\ref{eq17}) can be made by use of the gradient 
formula. A derivation of this formula can be found for instance in
chapter 5 of \cite{edm74}, from which one has
\begin{eqnarray}
\lefteqn{ \bm{\nabla}(\phi(r) Y_{\ell m}(\Omega_r))=}
                                  \nonumber \\ &&
-\left( \frac{\ell+1}{2\ell+1}\right)^{1/2}
\left(\frac{d}{dr}-\frac{\ell}{r} \right) \phi(r) \bm{Y}_{\ell,\ell+1,m}(\Omega) +
 \nonumber \\ & &
\left( \frac{\ell}{2\ell+1}\right)^{1/2}
\left(\frac{d}{dr}+\frac{\ell+1}{r} \right) \phi(r) \bm{Y}_{\ell,\ell-1,m}(\Omega),
\label{ap2}
\end{eqnarray}
where
\begin{equation}
\bm{Y}_{j,\ell,m}(\Omega)=\sum_{m,q} Y_{\ell,m}(\Omega) \langle \ell,m; 1,q |j,m\rangle
                           \bm{e}_q
\end{equation}
with $\bm{e}_0=\bm{e}_z$, and $\bm{e}_{\pm1}=\mp(\bm{e}_x \pm i
\bm{e}_y)/\sqrt{2}$.

Use of this expression permits to rewrite Eq.(\ref{eq17}) as:
\begin{small}
\begin{eqnarray}
 & & [[O_{\mu}^{\lambda\dagger},T],O_{\mu}^{\lambda}] = 
 \frac{\hbar^2}{m_N} \sum_{i,k=1}^3 
(c_i)^{\lambda} (c_k)^{\lambda} z_i z_k
 \sum_{\ell=1}^{\lambda} \sum_{m,n} \sqrt{2\lambda+1} \nonumber \\ & & 
 \left ( \begin{array}{ccc} \ell&\lambda-\ell&\lambda \\ m&n&-\mu
\end{array}\right) \sqrt{\frac{4\pi (2\lambda+1)!}{(2\ell+1)! 
 (2\lambda-2\ell+1)!}} \nonumber \\ & &  x_k^{\lambda-\ell} 
 Y_{\lambda-\ell, n}(\Omega_{x_k}) 
 (\sin \varphi_{ik})^{\lambda-\ell} (\cos \varphi_{ik})^l 
  y_k^{\lambda+\ell-2}
 \label{eq18} \\ & & (2\lambda+1) (2\ell+1) \sqrt{\ell\lambda}
 \sum_{q \eta \nu} (-1)^q Y_{\lambda-1, \eta} 
 (\Omega_{y_k}) Y_{\ell-1, \nu} (\Omega_{y_k}) \nonumber \\ & & 
(-1)^{\lambda-\mu+\ell+m} \left ( \begin{array}{ccc} \lambda-1&1&\lambda 
 \\ \eta&q&\mu
\end{array}\right) \left ( \begin{array}{ccc} \ell-1&1&\ell \\ \nu&-q&-m
\end{array}\right) \nonumber \; .
\end{eqnarray}
\end{small}

Writing now the two spherical harmonics in terms of a single one,
and summing up three of the 3-j symbols \cite{rot59} one gets:
\begin{small}
\begin{eqnarray}
 & & \hspace{-5mm} [[O_{\mu}^{\lambda\dagger},T],O_{\mu}^{\lambda}] = \frac{\hbar^2}{m_N} \sum_{i,k=1}^3 
(c_i)^{\lambda} (c_k)^{\lambda} z_i z_k \sum_{\ell=1}^{\lambda} \sum_{m,n} \sqrt{2\lambda+1} 
\\ & & \hspace{-5mm} \left ( \begin{array}{ccc} \ell&\lambda-\ell&\lambda \\ m&n&-\mu
\end{array}\right) \sqrt{\frac{(2\lambda+1)!}{(2\ell+1)! (2\lambda-2\ell+1)!}} x_k^{\lambda-\ell} Y_{\lambda-\ell, n}(\Omega_{x_k}) \nonumber \\ & & \hspace{-5mm} (\sin \varphi_{ik})^{\lambda-\ell} (\cos \varphi_{ik})^l \sqrt{\ell} (2\ell+1) \sqrt{\lambda} (2\lambda+1) y_k^{\lambda+\ell-2} \nonumber \\
 & & \hspace{-5mm} \sum_{\Lambda} (-1)^{1-\mu+m-n} \sqrt{(2\lambda-1)(2\ell-1)(2\Lambda+1)} Y_{\Lambda, n}^{*}(\Omega_{y_k}) \nonumber \\
 & & \hspace{-5mm} \left ( \begin{array}{ccc} \lambda-1&\ell-1&\Lambda \\ 0&0&0
\end{array}\right) \left ( \begin{array}{ccc} \lambda&\ell&\Lambda \\ \mu&-m&-n
\end{array}\right) \left \{ \begin{array}{ccc} \lambda&\ell&\Lambda \\ \ell-1&\lambda-1&1
\end{array}\right\} \nonumber \;,
\end{eqnarray}
\end{small}
\noindent
which after summation over $m$ and $\mu$ (the summation over $\mu$
comes from (\ref{eq12})), leads to Eq.(\ref{eq19}) for
the contribution of the kinetic energy operator to the energy-weighted
sum rule $S_1^{(T)}$

We close the appendix giving analytical expressions for the two matrix elements
entering in Eq.(\ref{eq30b}) for the particular case of particles without spin.
The expressions are obtained performing analytically the integrals over
$\Omega_{x_i}$ and $\Omega_{y_i}$:
\begin{small}
\begin{eqnarray}
 & & \langle {\cal Y}_{\ell_{x_i} \ell_{y_i}}^{K L_0 M_0}(\Omega_i)|
  O_\mu^{\lambda \dagger} |{\cal Y}_{\ell''_{x_i} \ell''_{y_i}}^{K'' L'' M''}(\Omega_i)\rangle= 
(-1)^{\mu+L_0+M_0+\ell''_{y_i}-\ell''_{x_i}}
\nonumber \\ & & 
\sqrt{\frac{(2L''+1)(2L_0+1)(2\ell''_{x_i}+1)(2\ell_{x_i}+1)(2\ell''_{y_i}+1)(2\ell_{y_i}+1)}{4 \pi}} 
\nonumber \\ & & 
\left ( \begin{array}{ccc} L''&L_0&\lambda \\ -M''&M_0&\mu \end{array}\right) 
N_{K''}^{\ell''_{x_i} \ell''_{y_i}} N_{K}^{\ell_{x_i} \ell_{y_i}}
\sum_{\ell=0}^{\lambda} 
\sqrt{\frac{2 \lambda!}{2 \ell! (2 \lambda-2 \ell)!}} 
\nonumber \\ & & 
\left ( \begin{array}{ccc} \ell''_{y_i}&\ell&\ell_{y_i} \\ 0&0&0 \end{array}\right) 
\left( \begin{array}{ccc} \ell''_{x_i}&\lambda-\ell&\ell_{x_i} \\ 0&0&0 \end{array}\right) 
\left\{ \begin{array}{ccc} L''&L_0&\lambda \\ \ell''_{x_i}&\ell_{x_i}&\lambda-\ell \\ \ell''_{y_i}&\ell_{y_i}&\ell
                  \end{array}\right \}
  \nonumber \\ & &
\rho^{\lambda} \sum_{k=1}^3 z_k (c_k)^{\lambda}
\int_0^{\pi/2} d\alpha_i (\sin \alpha_i)^{\ell''_{x_i}+\ell_{x_i}+2} 
(\cos \alpha_i)^{\ell''_{y_i}+\ell_{y_i}+2} 
\nonumber \\ & & 
P_{\nu''_i}^{(\ell''_{x_i}+\frac{1}{2},\ell''_{y_i}+\frac{1}{2})}(\cos 2 \alpha_i) 
P_{\nu_i}^{(\ell_{x_i}+\frac{1}{2},\ell_{y_i}+\frac{1}{2})}(\cos 2 \alpha_i) 
\nonumber \\ & & 
(\sin \varphi_{ki})^{\lambda-\ell} (\cos \varphi_{ki})^{\ell} 
(\sin \alpha_i)^{\lambda-\ell} (\cos \alpha_i)^\ell \; ,
\label{ap6}
\end{eqnarray}
\end{small}
where $N_{K}^{\ell_{x} \ell_{y}}$ is the normalization constant of the
hyperspherical harmonic ${\cal Y}_{\ell_{x} \ell_{y}}^{K L M}(\Omega)$,
whose precise form can be found for instance in \cite{nie01}.

The expression for the second matrix element in Eq.(\ref{eq30b}) is
identical to Eq.(\ref{ap6}) but with the function
$\left(V_{i}^{(\ell''_{x_i})}(\rho \sin \alpha_i)-
V_{i}^{(\ell'_{x_i})}(\rho \sin \alpha_i)\right)$ included as a factor
in the integrand, and with primes on the quantum numbers $K$,
$\ell_{x_i}$ and $\ell_{y_i}$.


\begin{thebibliography}{99}

\bibitem{sha74} A. de Shalit and H. Feshbach, Theoretical Nuclear Physics 
(Wiley, New York, 1974).

\bibitem{boh75} A. Bohr and B.R. Mottelson, Nuclear Structure, vol II,
(Benjamin, Reading, Massachusetts, 1975).

\bibitem{boh79} O. Bohigas, A. M. Lane and J. Martorell,
Phys. Rep. {\bf 51}, 267 (1979).

\bibitem{lip89} E. Lipparini and S. Stringari, 
Phys. Rep. {\bf 175}, 103 (1989).

\bibitem{jen04} A.S.~Jensen, K. Riisager, D.V.~Fedorov and E. Garrido,  
Rev. Mod. Phys. {\bf 76}, 215 (2004).

\bibitem{sag90} H. Sagawa and M. Honma, Phys. Lett. {\bf B251}, 17 (1990).

\bibitem{suz03}  T. Suzuki, H. Sagawa and K. Hagino,
 Phys. Rev. {\bf C68}, 014317 (2003).

\bibitem{myo01} T. Myo, K. Kato, S. Aoyama, and K. Ikeda,
Phys.Rev. {\bf C 63}, 054313 (2001).

\bibitem{des06} P. Descouvemont, E. Tursunov and D. Baye, 
Nucl. Phys. {\bf A 765}, 370 (2006).

\bibitem{alv07} R. Alvarez-Rodriguez, E. Garrido,  A.S. Jensen,  
D.V. Fedorov, and H.O.U. Fynbo, Eur.Phys.J. {\bf A 31}, 303 (2007). 

\bibitem{kan07} Y. Kanada-Enyo, Prog. Theor. Phys. {\bf 117}, 655 (2007). 

\bibitem{mas07}  H. Masui, K. Kato, and K. Ikeda,
Phys. Rev. {\bf C 75}, 034316 (2007).

\bibitem{nie01} E. Nielsen, D.V. Fedorov, A.S. Jensen, and E. Garrido,
Phys. Rep. {\bf 347}, 373 (2001).

\bibitem{gar02} E. Garrido, D.V. Fedorov and A.S. Jensen, 
Nucl. Phys. {\bf A 700}, 117 (2002).

\bibitem{gar04} E. Garrido, D.V. Fedorov and A.S. Jensen, 
Nucl. Phys. {\bf A 733}, 85 (2004).

\bibitem{rot59} M. Rotenberg, R. Bivins, N. Metropolis, and J.K. Wooten,
\textit{The 3-j and 6-j symbols}, (Technology Press, MIT, Cambridge MA, 1959)

\bibitem{gar99} E. Garrido and E. Moya de Guerra, 
Nucl. Phys. {\bf A650}, 387 (1999).

\bibitem{aum99} T. Aumann {\it et al.}, Phys. Rev. {\bf C59}, 1252 (1999).

\bibitem{dan98} B.V. Danilin, I.J. Thompson, J.S. Vaagen, and M.V. Zhukov,
Nucl. Phys. {\bf A632}, 383 (1998).

\bibitem{suz91} Y. Suzuki, Nucl. Phys. {\bf A528}, 395 (1991).

\bibitem{dan93} B.V. Danilin, M.V. Zhukov, J.S. Vaagen, and J.M. Bang, Phys. Lett. {\bf B302}, 129 (1993).

\bibitem{edm74} A.R. Edmonds, \textit{Angular Momentum in Quantum Mechanics},
(Princeton University Press, 1974).

\end{thebibliography}
\end{document}